\documentclass[
 reprint,
amsmath,amssymb,
prl,
]{revtex4-1}

\usepackage{graphicx}
\usepackage{amssymb}
\usepackage{booktabs}
\usepackage{xspace}

\newcommand{\ccb}{\ensuremath{c\overline{c}}}
\newcommand{\bbb}{\ensuremath{b\overline{b}}}
\newcommand{\ttb}{\ensuremath{t\overline{t}}\xspace}
\newcommand{\hh}{\ensuremath{hh}\xspace}
\newcommand{\ptdijet}{\ensuremath{p_{\mathrm T}^{\mathrm{dijet}}}}

\begin{document}

\preprint{APS/123-QED}

\title{Boosted  \hh$\rightarrow$~\bbb\bbb: a new topology in searches for TeV-scale resonances at the LHC}

\author{Ben Cooper}
\author{Nikos Konstantinidis}%
\author{Luke Lambourne}%
\author{David Wardrope}%
\affiliation{%
Department of Physics and Astronomy, University College London, Gower Street, London WC1E 6BT, United Kingdom
}

\begin{abstract}
It is widely believed that fully hadronic final states are not competitive in searches for new physics at the Large Hadron Collider 
due to the overwhelming QCD backgrounds.
In this letter, we present a particle-level study of the topology arising when a
TeV-scale resonance decays to two Higgs bosons and these subsequently decay to \bbb, leading
to two back-to-back boosted dijet systems. We show that selecting events with this topology
dramatically reduces all 
backgrounds, thus enabling very competitive searches for new physics in a variety of models.  For a resonance with mass 1\,TeV and width around 60\,GeV,
we find that ATLAS or CMS could have a sensitivity to a $\sigma \times
BR$ as small as a few fb with the LHC data collected in 2012.
These conclusions are also relevant to
the boosted $Zh\rightarrow$~\bbb\bbb\ and $ZZ\rightarrow$~\bbb\bbb\ final states, which would further increase the
potential sensitivity to new physics as well as to Standard Model processes like longitudinal vector boson scattering.

\end{abstract}

\keywords{LHC \sep boosted $b\overline{b}$ topologies \sep TeV-scale resonances \sep vector boson scattering \sep 2 Higgs doublet models \sep graviton}
\pacs{12.60.nz and 12.60.cn and 12.60.fr and 12.60.jv}

\maketitle

Since the start of collision data-taking in 2009, the LHC experiments have performed numerous analyses searching for high mass resonances at the TeV scale.
With few exceptions~\cite{CMS:2012yf, Chatrchyan:2013qha, Chatrchyan:2012ypy, ATLAS-CONF-2012-148}, 
such searches have relied on final states containing leptons (electrons or muons), for
example: $Z^{\prime}\rightarrow \ell^+\ell^-$~\cite{Aad:2012hf, Chatrchyan:2012oaa}; 
resonant diboson production ($WW$, $WZ$, $ZZ$)~\cite{Chatrchyan:2012rva, Aad:2013wxa, ATLAS-CONF-2012-150}, where at least one vector boson decays
to leptons; or searches for resonances decaying to \ttb~\cite{Chatrchyan:2012yca, Aad:2013nca, Collaboration:1543917}, 
where one or both of the top quarks decay semi-leptonically. The main reasons for this choice 
are that the presence of leptons helps to drastically reduce the QCD backgrounds, and that such final states are relatively easy to trigger on in ATLAS and CMS. However, the low branching ratios of the leptonic decays of the vector bosons, particularly of 
the $Z$, mean that the above searches are not sensitive to a large fraction of the production cross section of possible new resonances. In 
addition, resonances with masses at the TeV scale generally have a large natural width (typically tens of GeV or more), so
the good experimental mass resolution offered by the leptonic channels does not improve the search sensitivity as much as for low mass resonances, where strict mass selection criteria are possible due to small widths. Finally, in many theoretical 
models~\cite{Randall:1999ee, Coleppa:2013dya, Pruna:2013bma}, high mass resonances often decay with a sizeable branching
ratio to a pair of Higgs bosons and since the Standard Model (SM) predicts the dominant decay of the Higgs to be to \bbb, trying to look for final states involving leptons (or photons) from the Higgs decays leads 
to a significant effective reduction of the signal cross section.

Motivated by these arguments and by the discovery of the Higgs boson~\cite{Aad:2012tfa, Chatrchyan:2012ufa}, $h$, which appears to be consistent with the SM expectations, we have performed a
particle-level analysis to evaluate the LHC sensitivity to TeV-scale resonances decaying to \hh and subsequently to \bbb\bbb. For resonances of this mass scale, the two Higgs bosons will have high transverse momenta, resulting in two highly boosted, back-to-back \bbb\ dijet systems.
This topology has several advantages: (i) requiring four $b$-tagged
jets paired into two boosted dijets is a very powerful way to drastically reduce backgrounds, in particular QCD; 
(ii) there is negligible ambiguity in pairing the four $b$-jets to
correctly reconstruct the Higgs decays; and (iii) due to the high boost, the four jets will have high enough transverse momenta
for such events to be selected with high efficiency at the first level trigger of ATLAS and CMS, with efficient High Level triggering possible through online $b$-tagging. The same topology will also arise for any high mass resonance decaying to \bbb\bbb~via $Zh$ or $ZZ$. In the following, 
we focus on the $hh\rightarrow$~\bbb\bbb\ 
final state, but the conclusions of this study are also relevant for the $Zh\rightarrow$~\bbb\bbb\ and $ZZ\rightarrow$~\bbb\bbb\ final states. 

The signal used as a benchmark in this study is a Randall-Sundrum~\cite{Randall:1999ee} Kaluza-Klein (KK) graviton, $G_{\rm KK}$, decaying
to $hh$, but several other models beyond the SM lead to this topology, 
for example $H\rightarrow hh$ in two-Higgs-doublet models (2HDM)~\cite{Coleppa:2013dya}, or in singlet extensions to the SM Higgs 
sector~\cite{Pruna:2013bma}. 
Furthermore, this topology will also
arise in: (i) decays of the pseudoscalar Higgs boson of 2HDM, $A\rightarrow Zh$; and (ii) longitudinal $ZZ$ 
boson scattering at the TeV scale. The latter offers an alternative channel with which to measure the vector boson scattering cross section
and to search for possible resonances that may modify this.

The KK graviton signal is generated using
MadGraph~\cite{Alwall:2011uj} based on the scenario proposed in
\cite{PhysRevD.76.036006, 1126-6708-2007-09-013} with $k/\bar{M}_{\rm Pl} = 1$. The CTEQ6L1 parton density
functions (PDFs)~\cite{Stump:2003yu} were used and MadGraph was
interfaced to Pythia 8.170~\cite{Sjostrand:2007gs} for parton showering, hadronization and underlying event simulation. Table~\ref{tab:signal} shows the
width and predicted production cross section for various graviton
masses, $m_G$.

\begin{table}[htb]
\begin{center}
\begin{tabular}{ccccccccc}
\hline\toprule
Graviton Mass & $\sigma(pp\rightarrow G_{\rm KK} \rightarrow hh \rightarrow\bbb\bbb)$ & $\Gamma$ \\
{[}GeV{]} & [fb] & [GeV] \\
\hline
500 & 329 & 18.6 \\
700 & 72.7 & 33.9 \\
900 & 18.6 & 48.6 \\
1100 & 5.51 & 62.7 \\
1300 & 1.82 & 76.5 \\
1500 & 0.65 & 90.0 \\
\hline
\end{tabular}
\caption[graviton properties.]{Mass, cross section and width of the KK
  graviton signal. }
\label{tab:signal}
\end{center}
\end{table}

The event selection starts by requiring at least four $b$-tagged jets
with $p_{\mathrm T}>40$\,GeV and $|\eta|<2.5$. 
Jets are formed using the anti-$k_{\mathrm T}$ algorithm
\cite{Cacciari:2008gp} with radius parameter $R=0.4$\,, implemented in Fastjet~\cite{fastjet}. 
In order to estimate the effect of $b$-tagging, 
jets are labeled as $b$-jets, $c$-jets or light jets depending on the flavor of
partons within  $\Delta R<0.3$ of the jet axis. If a $b$-quark is found, the jet is labeled a $b$-jet, 
otherwise if a $c$-quark is found the jet is labeled a $c$-jet. If neither a $b$-quark nor a $c$-quark is found, then the jet is classified as a light jet.
We then apply $b$-tagging efficiency factors, based on the published ATLAS
and CMS $b$-tagging performance~\cite{Chatrchyan:2012jua, ATLAS-CONF-2012-097}: 70\% for 
$b$-labeled jets, 20\% for $c$-labeled jets (``rejection factor'' 5) and 1\% for light-labeled jets (``rejection factor'' 100).
Dijets are then formed, requiring $\ptdijet>200$\,GeV and
$\Delta R_{\mathrm{dijet}} = \sqrt{\Delta \phi^2 + \Delta \eta^2} <1.2$, where $\Delta \phi$ is the angular separation between jets in the
plane transverse to the beam line and $\Delta \eta $ is the difference in pseudorapidity between the two jets. 
Finally, we require a dijet invariant mass, $m_{\mathrm{dijet}}$, consistent with $h$ decay. 
In the extremely rare cases where more than two dijets satisfy all the above requirements, only the two dijets with the highest \ptdijet\ are considered.

\begin{figure}[hbt]
   \begin{center}
    \includegraphics[width=0.45\textwidth]{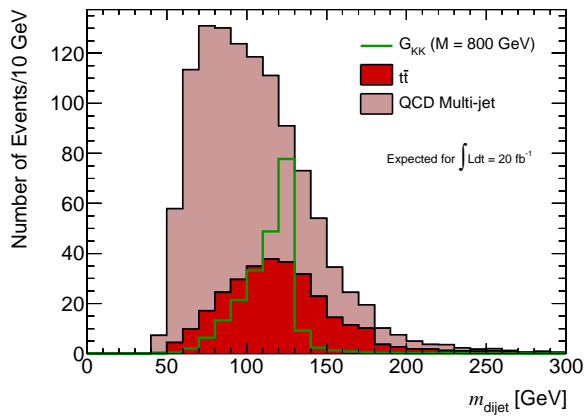}
 \caption {The individual dijet mass distributions in signal and
   background (stacked).
  \label{fig:mbbSignal}} 
   \end{center}
\end{figure}

The dijet mass distributions in the signal and background
are shown in Fig.~\ref{fig:mbbSignal} for $m_{\mathrm h}=125$\,GeV. It can be seen that the Higgs mass peak in the signal is shifted to lower masses. 
This is because, in this particle-level study, the reconstructed jets do not include neutrinos and muons or any out-of-cone corrections. 
Hence, we have chosen the Higgs mass window to
be $100<m_{\mathrm{dijet}}<130$\,GeV. 
This study has not considered detector resolution effects that will smear the jet $p_{\mathrm T}$ measurements and broaden the
four-jet invariant mass, $m_{4b}$. However, for resonances with natural width some tens of GeV or more, the detector resolution is not expected to increase 
the width of the observed peak significantly. Moreover, it is possible to take advantage of the known mass of the two Higgs bosons and perform a
kinematic fit to determine $m_{4b}$, which will largely remove the
detector resolution effects. The impact of additional pile-up interactions
has also not been considered in this study, since for
jets with $p_{\mathrm T}>40$\,GeV and $|\eta|<2.5$ both ATLAS and CMS
have demonstrated that pile-up effects can be strongly mitigated~\cite{ATLAS-CONF-2013-083, CMS-PAS-JME-13-005}.

\begin{figure}[hbt]
   \begin{center}
    \includegraphics[width=0.45\textwidth]{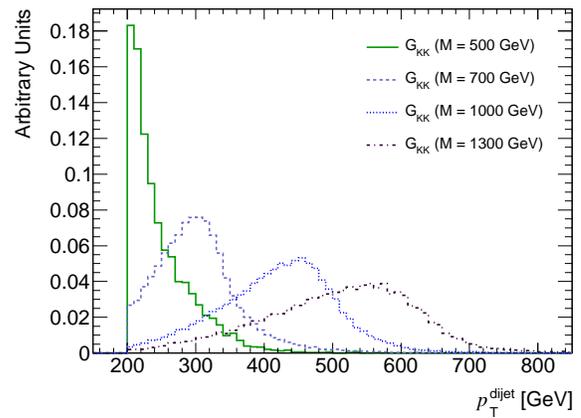}
 \caption {The $\ptdijet$ distribution for different graviton masses.
  \label{fig:dijetpt}}
   \end{center}
\end{figure}
The \ptdijet\ distribution is shown in Fig.~\ref{fig:dijetpt} for various signal masses. It can be seen that for $m_G=500$\,GeV, 
the $p_{\mathrm T}^{\mathrm{dijet}} > 200$\,GeV requirement has a visible impact on the signal acceptance and would need to be optimized to achieve the best possible sensitivity 
to masses around 700\,GeV and below. Similarly, for higher signal masses,
the optimal \ptdijet\ requirement would likely be higher than 200\,GeV.

\begin{figure}[hbt]
   \begin{center}
    \includegraphics[width=0.45\textwidth]{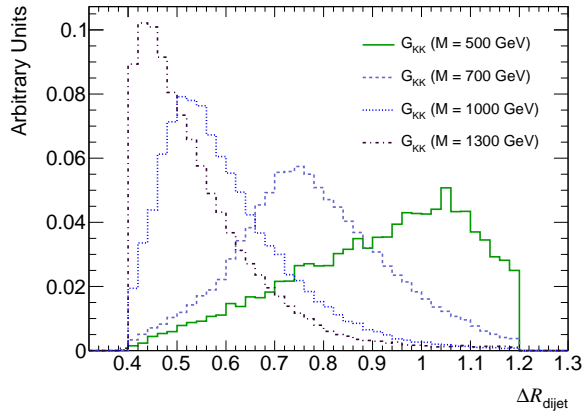}
 \caption {The $\Delta R_{\mathrm{dijet}}$ distribution for different graviton masses.
  \label{fig:deltaR}}
   \end{center}
\end{figure}

The $\Delta R_{\mathrm{dijet}}$ distribution in the dijet systems is
shown in Fig.~\ref{fig:deltaR} for various signal masses. It can be
seen that for high graviton masses the two $b$-jets from the $h$ decay tend to get closer and closer, hence with increasing likelihood merge into a single jet, leading to selection inefficiencies. 
This inefficiency can be dealt with in a number of ways: (i) by
forming narrower jets in the search for higher mass signals; (ii) by
identifying two separate $b$-hadron decay vertices within single,
energetic jets, as in~\cite{Aad:2013fk}; 
or (iii) by using jet substructure 
techniques~\cite{Butterworth:2008iy, Gouzevitch:2013qca} and applying $b$-tagging to the individual sub-jets. These alternatives should be explored in order 
to extend the sensitivity of this search to as high a resonance mass as possible. At the other end of 
the mass spectrum, the requirement $\Delta R_{\mathrm{dijet}} <1.2$ may be an
impediment for $m_{G} <$ 700\,GeV and should be revisited for extending the sensitivity of
the search to lower resonance masses.

\begin{figure}[hbt]
   \begin{center}
    \includegraphics[width=0.45\textwidth]{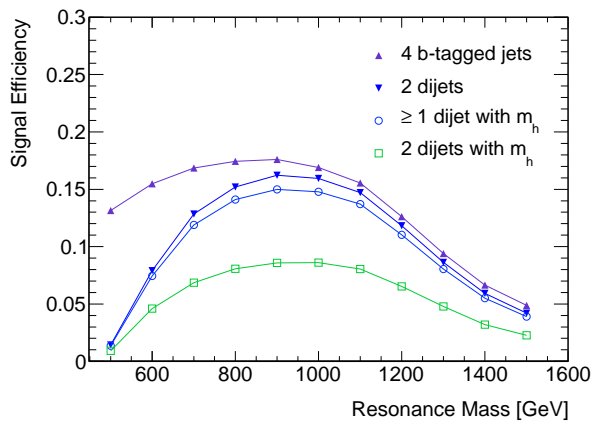}
 \caption{
 The evolution of signal efficiency at each step of the analysis. 
  \label{fig:signaleff}}
   \end{center}
\end{figure}

The signal efficiency through the cut flow of the above selection is shown in Fig.~\ref{fig:signaleff}. This confirms the observations made above about 
signal efficiency losses at the low and high ends of the resonance mass range considered in this study. The biggest efficiency drop
comes from the $b$-tagging requirement, approximately ($0.7^4\approx0.24$).

The backgrounds to this search are the irreducible pp\,$\rightarrow$\,\bbb\bbb\ and those reducible backgrounds resulting from the false classification of jets as $b$-jets, such as \ttb or pp\,$\rightarrow$\,\bbb\ccb. Other backgrounds, such as diboson production or $Z+$jets events, were found to be negligible due to the requirement that the dijet masses fall within
the $h$ mass range.

The pp\,$\rightarrow$\,\bbb\bbb\ and other QCD multi-jet backgrounds
have been simulated using Sherpa 1.4.3~\cite{Gleisberg:2008ta}, with
which we generate events based on tree-level matrix elements with four partons in the final state. After $b$-tagging, the QCD multi-jet background is dominated by pp\,$\rightarrow$\,\bbb\bbb, with a smaller contribution from mistagged pp\,$\rightarrow$\,\bbb\ccb\ events. 
The accuracy of the simulation of pp\,$\rightarrow$\,\bbb\bbb\ by Sherpa is verified by reproducing a total cross-section approximately equal to the leading order prediction from \cite{PhysRevLett.107.102002} and \cite{Bevilacqua:2013taa}. 
The uncertainty due to missing higher order matrix element terms in
pp\,$\rightarrow$\,\bbb\bbb\ and pp\,$\rightarrow$\,\bbb\ccb\ is
estimated by varying the renormalization and factorization scales by
factors of 1/2 and 2 from their nominal values of
$\frac{1}{4}\sqrt{\sum\limits_{i} p_{T,i}^{2}}$\,. 
It was found in \cite{PhysRevLett.107.102002} that this variation
covers the NLO prediction. Consequently, we are confident that our
estimate of this background is reliable within its uncertainty.

The \ttb\ background was simulated using Pythia 8.170, with the cross section scaled to the average cross
section measured by ATLAS and CMS at 8\,TeV~\cite{ATLAS-CONF-2012-149,
  CMS-PAS-TOP-12-027}.
We find that the flavor composition of the four-jet system in the \ttb\ events passing the full selection is predominantly $bc-bc$, which occurs when both $t$-quarks decay 
hadronically, as $t\rightarrow bW \rightarrow bcs$, and the $b$ and
$c$ jets end up nearby. The uncertainty on the measured cross section
is propagated as the uncertainty on the \ttb\ background estimate.

\begin{table}[htb]
\begin{center}
\begin{tabular}{lcccccc}
\hline
Requirement 	& $\rm{G_{KK} (M = 800\,GeV)}$ & QCD  		& \ttb\ 		\\ 
\hline
4 $b$-tagged jets & 126 & 19700 & 3590\\
2 dijets & 109 & 414 & 151\\
$\geq 1$ dijet with $m_{h}$ & 102 & 183 & 89\\
2 dijets with $m_{h}$ & 58 & $28_{-11}^{+20}$  & 21 $\pm$ 3 \\
\hline
\end{tabular}
\caption[Expected yields.]{Event selection, showing the expected signal and background yields
  for $\int\mathcal{L}dt = 20 \rm fb^{-1}$ at $\sqrt{s}=8$\,TeV. 
}
\label{tab:cutflow}
\end{center}
\end{table}
Table~\ref{tab:cutflow} shows the expected yields for a graviton signal with mass 800\,GeV and the QCD and \ttb\
backgrounds, along with their uncertainties.
Thanks to the powerful background rejection from the event topology requirements, the signal to background ratio is approximately 1 for this benchmark model, demonstrating the sensitivity of this analysis
to signals with very low cross section.

\begin{figure}[hbt]
   \begin{center}
    \includegraphics[width=0.45\textwidth]{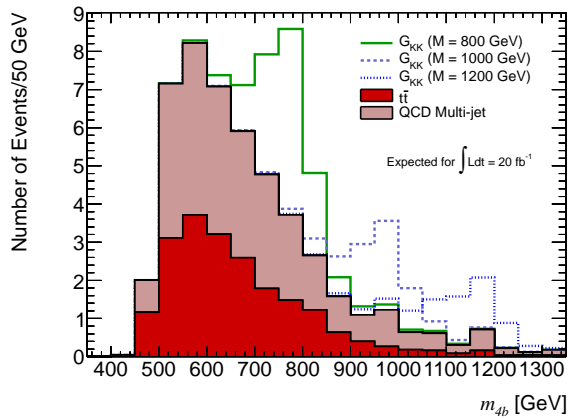}
 \caption{
 The stacked $m_{4b}$ distribution for various signal masses and for the
 background. The signal production cross-sections are normalized to
 those required for $3\sigma$ evidence in 20\,fb$^{-1}$ of pp collisions at $\sqrt{s}=8$\,TeV.
  \label{fig:mbbbb}}
   \end{center}
\end{figure}

\begin{figure}[hbt]
   \begin{center}
    \includegraphics[width=0.45\textwidth]{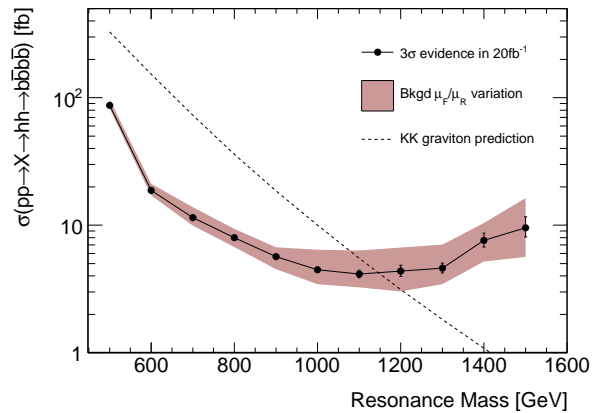}
 \caption{
 The signal cross section required in 20\,fb$^{-1}$ of pp collisions at $\sqrt{s}=8$\,TeV for achieving a $3\sigma$ evidence 
 of $pp\rightarrow X \rightarrow \hh \rightarrow$ \bbb\bbb. 
The error bars indicate the statistical uncertainties, and the shaded band the QCD background renormalization and factorization
 scale uncertainties. 
The $G_{KK}\rightarrow\hh\rightarrow\bbb\bbb$ cross section 
 from Table~\ref{tab:signal} is displayed as the dashed curve.
   \label{fig:sensitivity}}
   \end{center}
\end{figure}

The $m_{4b}$ distribution for signal and background is shown in
Fig.~\ref{fig:mbbbb}. To estimate the
boosted \bbb\bbb\ search sensitivity for
20\,fb$^{-1}$ of pp collisions at $\sqrt{s}=8$\,TeV, as a
function of the mass of a graviton-like resonance, we define mass
windows around the resonance mass, $m_X$, as $[m_X-100, m_X+50]$ (in GeV). We then determine the signal efficiency and expected number of background 
events, $N_{bkg}$, within these windows, and calculate the cross section $\sigma(pp\rightarrow X \rightarrow hh \rightarrow $ \bbb\bbb) that would
give $3\times \sqrt{N_{bkg}}$ signal events. These estimates of the 3$\sigma$ observation sensitivity are shown in Fig.~\ref{fig:sensitivity}\,. It can be seen that around 1\,TeV the analysis
achieves the best sensitivity, to cross-sections down to a few fb. At lower and higher masses the loss of signal acceptance mentioned above 
decreases the sensitivity, but as already commented, there is scope for optimization that could bring the sensitivity to O(10fb) in these regions, 
and extend the reach much beyond the mass range considered here. 
Although experimental systematics that would worsen the sensitivity have not been considered, large gains in sensitivity are possible through an overall optimization of the 
analysis. These results should apply equally well to any model that predicts a high mass resonance decaying to \hh,
provided that its natural width is similar to that of the KK graviton used in this study. For the KK graviton model used here, Fig.~\ref{fig:sensitivity} shows that this analysis would give more than
$3\sigma$ sensitivity up to the TeV scale.

For many new physics models~\cite{Coleppa:2013dya, Pruna:2013bma}, where the decay of the intermediate
resonance to anything other than $hh$ can be strongly suppressed, the results in Fig.~\ref{fig:sensitivity}
represent a unique sensitivity. For others, the sensitivities shown in Fig.~\ref{fig:sensitivity} could be
surpassed by combining the $hh \rightarrow $ \bbb\bbb\ search with other channels and final
states. For example, in the signal model used in this letter the KK graviton
decays to both $hh$ and $ZZ$ with branching fractions around 10\%.  A KK graviton search in
the \bbb\bbb\ final state combining both $ZZ$ and $hh$ channels would increase considerably the sensitivity to such a
signal, and could be further combined with the already explored $WW$ and $ZZ$
search channels that don't involve $b$-quarks in the final
state~\cite{Chatrchyan:2012ypy,Chatrchyan:2012rva, Aad:2013wxa, ATLAS-CONF-2012-150}.  

We conclude that the final state topology of two, boosted, $b$-tagged dijet systems shows great promise for increasing the LHC sensitivity to TeV-scale 
resonances decaying into a pair of electroweak-scale bosons, such as $hh$,
$Zh$ or $ZZ$. It combines
low background levels with good acceptance to various signal processes, making it 
a powerful channel in which to search for new physics, both in the current LHC data and in the higher energy running from 2015 onwards.

\end{document}